\documentclass[prl,aps,twocolumn, floatfix, superscriptaddress,dvipdfmx]{revtex4}
\usepackage{amsmath,bm,graphicx}
\usepackage{amsmath}
\usepackage{amssymb}
\usepackage{amsfonts}
\begin{document}
\title{Sensitivity of Granular Force Chain Orientation to Disorder-induced Metastable Relaxation}

\author{N. Iikawa}
\affiliation{Department of Earth and Environmental Sciences, Nagoya University, Nagoya 464-8601, Japan}
\author{M. M. Bandi}
\affiliation{Collective Interactions Unit, OIST Graduate University, Onna, Okinawa 904-0496, Japan}
\author{H. Katsuragi~\footnote{Corresponding Author: katsurag@eps.nagoya-u.ac.jp}}
\affiliation{Department of Earth and Environmental Sciences, Nagoya University, Nagoya 464-8601, Japan}
\email[Corresponding Author: ]{katsurag@eps.nagoya-u.ac.jp}

\date{\today}

\begin{abstract}
A two-dimensional system of photoelastic disks subject to vertical tapping against gravity was experimentally monitored from ordered to disordered configurations by varying bidispersity. The packing fraction $\phi$, coordination number $Z$, and an appropriately defined force chain orientational order parameter $S$, all exhibit similar sharp transition with small increase in disorder. A measurable change in $S$, but not $\phi~\&~Z$, was detected under tapping. We find disorder-induced metastability does not show configurational relaxation, but can be detected via force chain reorientations.

\end{abstract}
\maketitle
Order-disorder transitions represent one of the most important concepts in statistical physics~\cite{Chaikin}. For example, magnetic and liquid-gas phase transitions are well understood through the concept of order-disorder transition. This concept finds applicability beyond these classical thermodynamic phase transitions in various soft matter problems including granular matter, liquid crystals etc. For instance, the crystalline (ordered) structure resulting from densely packed monodisperse grains gives way to a disordered structure when polydisperse grains are introduced in granular media; a type of athermal order-disorder transition. In such granular packs, configurational structure is supported by a set of grain-contact networks called force chains. Whereas force chain structure is related to the grain configuration, it also possesses an orientational degree of freedom, a situation similar to liquid crystal nematic phases where molecular positions are random but their orientations exhibit an ordered state \cite{DeGennes}; configurational structure and orientational order can be independent. Therefore, a quantification of the orientational order as well as structural order become necessary for the proper characterization of the order-disorder transition in granular matter. In this study, we experimentally measure structural and orientational parameters in granular order-disorder transition by using photoelastic materials~\cite{Oda1974}. As a consequence, we find the force chain orientational order parameter is much more sensitive than structural parameters, and can characterize the granular pack's relaxation under external perturbation owing to metastability arising purely from structural disorder, as opposed to mechanical sources such as friction.

\begin{figure*}
\begin{center}
\includegraphics[width = 6.74 in]{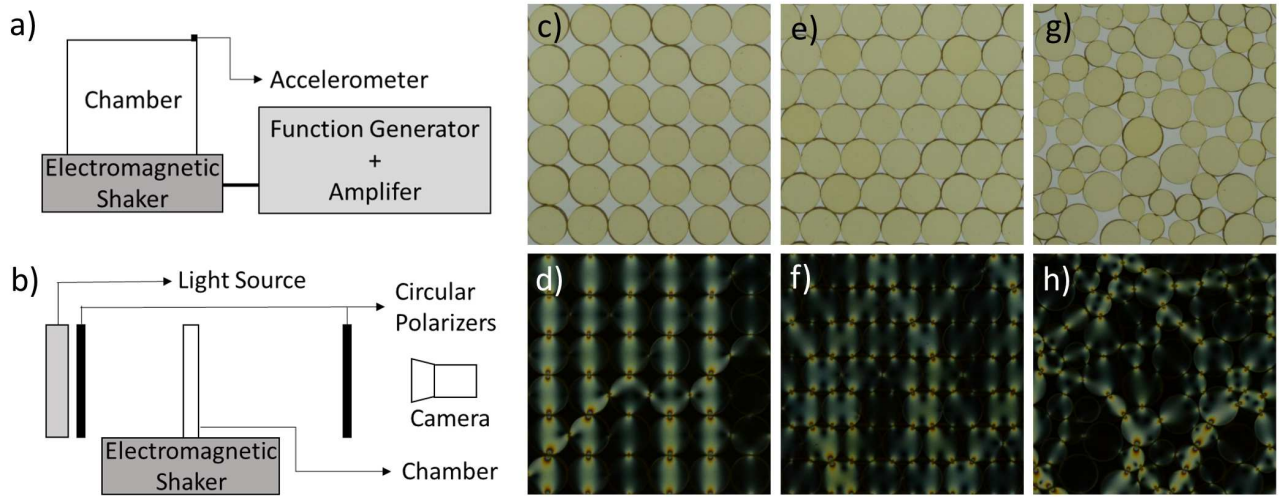}
\end{center}
\caption{((Color online) (a) Front and (b) side view schematics of the experimental setup. The disks were placed in a vertical chamber mounted on an electromagnetic shaker. An accelerometer was attached to top right corner of the chamber. A light source with a circular polarizer provided backlit illumination for the system. A second circular polarizer oriented in cross-polarization mode placed in front of the chamber was used for dark field imaging of photoelastic fringes. Representative images of disk configurations show: square lattice (c) bright and (d) dark, triangular lattice (e) bright and (f) dark, and random configuration with bidispersity ratio $R_L = R_S = 0.5$ (g) bright and (h) dark field images, respectively.}
\label{fig1}
\end{figure*}

The experimental setup (Fig.~\ref{fig1}a \& b) consisted of a bidispersed set of photoelastic disks placed within an acrylic chamber of dimensions $0.3 \times 0.3 \times 0.011$ m$^{3}$, held vertically on an electromagnetic shaker (EMIC 513-B/A) driven by a function generator via an amplifier (EMIC 374A/G). Disks were cut from a 0.01 m thick photoelastic sheet (Vishay Micromeasurements, PSM-4) with diameter $D_L = 0.015$ m (large) and $D_S = 0.01$ m (small). The granular pack's configurational disorder was controlled by varying bidispersity through the ratio of small to large disks and quantified using $R_S$ ($R_L$), the ratio of area occupied by small (large) disks to total area of the disks ($R_S + R_L = 1$). We varied the bidispersity ratio $R_S$ (alternatively $R_L$)  for 17 values as shown in Table~\ref{tabexpts}. Two types of ordered structures (square and triangular lattices) were produced for ordered states ($R_S=1$ or $R_L=1$). For disordered states, large and small disks were dispersed homogeneously to the extent possible, in their initial states. Then, both the disk configurations and force chain structures were optically captured by two modes: standard bright field imaging to obtain the pack's configurational information and circular polariscope dark field imaging to analyse force chain network~\cite{note2} (for details see \cite{Iikawa2015}). 

To investigate the stability of this initial packing state, we added tapping to the system by using a shaker. An accelerometer (EMIC 710-C) mounted on the chamber measured acceleration magnitude under tapping. Each initial configuration (tapping number~$\tau = 0$) was tapped 10 times, i.e. final state $\tau=10$. All experiments described here were performed at a dimensionless acceleration $\Gamma = \bm{a}/|\bm{g}| \simeq 10$, where $a$ is the maximum tapping acceleration and $|\bm{g}| = 9.8$ m/s$^2$. The corresponding amplitude $A$ normalized to the small disk diameter is $A/D_s \simeq 0.025$. Images were acquired after every tapping in addition to the initial state. Three experimental runs were performed in each experimental condition.

\begin{table*}
\begin{tabular}{|c|ccccccccccccccccc|}
\hline
Area Ratio ($R_L$) & 1 & 0.99 & 0.97 & 0.95 & 0.9 & 0.8 & 0.7 & 0.6 & 0.5 & 0.4 & 0.3 & 0.2 & 0.1 & 0.05 & 0.03 & 0.01 & 0\\
 Area Ratio ($R_S$) & 0 & 0.01 & 0.03 & 0.05 & 0.1 & 0.2 & 0.3 & 0.4 & 0.5 & 0.6 & 0.7 & 0.8 & 0.9 & 0.95 & 0.97 & 0.99 & 1\\
No. of large disks & 400 & 396 & 388 & 380 & 360 & 320 & 280 & 240 & 200 & 160 & 120 & 80 & 40 & 20 & 12 & 4 & 0\\
No. of small disks & 0 & 9 & 27 & 45 & 90 & 180 & 270 & 360 & 450 & 540 & 630 & 720 & 810 & 855 & 873 & 891 & 900\\
\hline
\end{tabular}
\caption{Area ratio $R_S$ ($R_L$) at which experiments were conducted. Three experimental runs were performed at each value. At $R_S = 0~\&~1$ ($R_L = 1~\&~0$), three sets each were collected for both triangular and square lattice configurations.}
\label{tabexpts}
\end{table*}

Quantities of interest extracted from image analysis include the packing fraction $\phi$~\cite{Iikawa2015}, coordination number $Z$~\cite{note1}, and force chain orientational order parameter~\cite{DeGennes} which characterizes the alignment of force chains to the gravitational director ($\hat{\bm{g}} \equiv \bm{g}/|\bm{g}|$). Each force chain running from one boundary to the other consists of smaller segments \cite{Iikawa2015}, where the $i$-th segment has length $l_i$ and angle $\theta_i$ relative to $\hat{\bm{g}}$. This permits definition of an orientational order parameter:
\begin{equation}
S = \left(\frac{2}{L}\sum_i l_i{\cos}^2\theta_i\right) - 1,
\label{S}
\end{equation}
where $L = \sum_i l_i$ is the total length of all force chains. In limiting cases, Eq.~(\ref{S}) yields $S = +1$ for $\theta = 0^{\circ}$ (force chains parallel to $\hat{\bm{g}}$) and $S = -1$ for $\theta = 90^{\circ}$ (force chains perpendicular to $\hat{\bm{g}}$). Randomly oriented force chains (isotropic orientation) or $\theta = 45^{\circ}$ yield $S = 0$. Visual inspection trivially distinguishes between perfect $\theta = 45^{\circ}$ and isotropic alignment for $S = 0$.

Figure~\ref{fig1} shows representative bright and dark field images of crystalline ($R_L=1$) square (Fig.~\ref{fig1}c \& d) and triangular (Fig.~\ref{fig1}e \& f) lattices, and a disordered configuration obtained at $R_S = R_L = 0.5$ (Fig.~\ref{fig1} g \& h), respectively. Despite the bright field images exhibiting perfect crystalline square (Fig.~\ref{fig1}c) and triangular (Fig.~\ref{fig1}e) lattice configurations, their corresponding dark field images (Fig.~\ref{fig1}d \& f) clearly show force chain defects due to minor disk imperfections. This force chain sensitivity is not well appreciated on account of their usual applicability to disordered granular packs (Fig.~\ref{fig1}g \& h) which yield highly heterogeneous structures. This simple empirical observation suggests that the force chain structures may be an extremely sensitive probe of structural information.

\begin{figure}
\begin{center}
\includegraphics[width = 3.37 in]{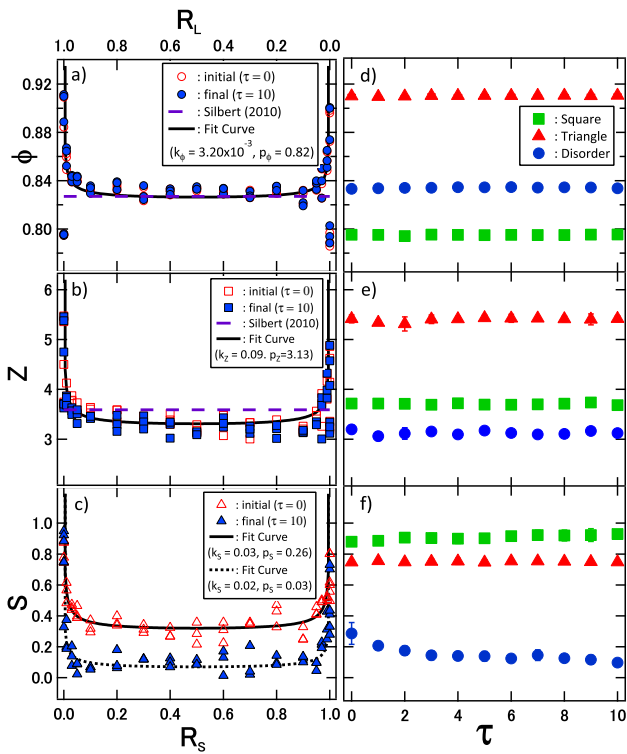}
\end{center}
\caption{(Color online) Left column: The initial ($\tau=0$; open symbols) and final ($\tau=10$; solid symbols) configurational values for (a) packing fraction $\phi$ (circles), (b) coordination number $Z$ (squares), and (c) force chain orientational order parameter $S$ (triangles) versus bidispersity metric $R_S$ ($R_L$). Curves are fits to initial (solid) and final (dotted, plotted only for $S$ with measurable change in values) configurational data following the functional form in Eq.~(\ref{fit}). Fit parameter values $k_X$ and $p_X$ are provided in respective legends. Right column: evolution of (d)~$\phi$, (e)~$Z$, and (f)~$S$ as a function of tapping number $\tau$ for crystalline square (green solid squares) and triangular (red solid triangles) as well as bidisperse disorder configurations (solid blue circles) for $R_S = R_L = 0.5$. Error bars represent variability observed over three experimental runs.}
\label{fig2}
\end{figure}

In the first set of experiments, we monitored the change in the structural parameters $\phi$, $Z$, and $S$ by varying $R_S$. In Fig.~\ref{fig2}, we plot $\phi$ (Fig.~\ref{fig2}a), $Z$ (Fig.~\ref{fig2}b), and $S$ (Fig.~\ref{fig2}c) versus $R_S$ ($R_L$) for the initial ($\tau = 0$, open symbols) and the final ($\tau = 10$, solid symbols) configurations. All three quantities exhibit a very similar, sharp structural transition over a short range in $R_S$ ($R_L$) from a completely ordered to an asymptotic disordered state. Small amount of bidispersity with $R_S=0.1$ (or $R_L=0.1$) is sufficient to achieve the asymptotic disordered sate (flat bottoms in Fig.~\ref{fig2}a, b, \& c). To characterize the degree of bidispersity, the dimensionless parameter $\mathcal{A}=\langle D \rangle^2/\langle D^2 \rangle$ was introduced in~\cite{Luding2001}, where $D$ is grain size and $\langle \cdot \rangle$ means average of the entire system. By this definition, $R_S=0.1$ corresponds to $\mathcal{A}=0.98$, a value consistent with numerically obtained criterion ($\mathcal{A}<0.99$) to avoid crystallization~\cite{Luding2001}. The result obtained here provides a practically useful criterion to avoid crystallization in experiments and simulations using bidispersed systems.

The asymptotic values for $\phi$ and $Z$ at disordered states are also in accord with expected values. Although the disordered granular packs with zero friction approach random close packing limit ($\phi \rightarrow 0.84~\&~Z \rightarrow 4$ in 2D), friction leads to reduction in both $\phi$ and $Z$. Owing to disks in our experiment having non-zero friction (friction coefficient $\mu \simeq 0.19$~\cite{Bandi2013}), the expected values for the jammed configurations in our experiments based on numerical results (see Table 1 in \cite{Silbert2010}) are $\phi \simeq 0.827~\&~Z \simeq 3.59$ (horizontal dashed lines in Fig.~\ref{fig2}a \& b, respectively), very close to our experimentally observed values. 

For the orientational order parameter $S$, we expect the crystalline square lattice configuration exhibits vertical force chains aligned along $\hat{\bm{g}}$, i.e., $S=1$. Whereas the ideal triangular lattice configuration ($\theta = 30^{\circ}$) results in $S=0.5$, the gravitational force will distort this orientational order to the gravitational direction $\hat{\bm{g}}$. As a result, $S$ becomes greater than 0.5 in actual triangular lattice configurations. In any case, values of $S$ for ordered states close to $R_S = 0~\&~1$ sharply approach $S \rightarrow 1$. However, the values of $S$ in disordered states ($0.1 \leq R_S \leq 0.9$) are roughly constant, similar to $\phi$ and $Z$ behavior. 

Fitting functions for these three plots (Fig.~\ref{fig2}a, b, \& c) follow the same empirical form:
\begin{equation}
X = \frac{k_X}{\sqrt{R_S R_L}} + p_X,
\label{fit}
\end{equation}
where $k_X$ and $p_X$ (see Fig.~\ref{fig2} legends for fit values) are the respective fitting parameters for $X = \phi,~Z,~{\text or}~S$. Since $R_S + R_L = 1$, Eq.~(\ref{fit}) follows the functional form $X \sim 1/\sqrt{R_S(1-R_S)}$ thus explaining the singular square-root behavior as one approaches $R_S~\&~R_L$ values close to ordered configurations. Please note however that none of the quantities diverge at the ordered state. Thus, Eq.~(\ref{fit}) is an empirical approximation. In addition, very low $\phi$ values (less than $0.8$) at $R_S=0$ \& $1$ coming from the square lattice are neglected here in the fitting. Starting with an ordered granular configuration $R_S = 0~{\text or}~ 1$, the replacement of small disks with large ones (while keeping total disk area constant) creates disordered clusters in the pack; a situation not dissimilar to structural transitions witnessed in well studied thermodynamic order-disorder transitions. For instance, crystalline melting characterised by vacancy creation \cite{Tabor, Kirkwood1950} causes non-symmetric 5- and/or 7-fold sites, whose increase and eventual formation of a spanning, disordered cluster results in a structurally disordered liquid \cite{Bernal1962}. Neither can the current experiments directly map to thermodynamic order-disorder transitions, nor is conclusive determination of square-root behavior in this structural order-disorder transition possible from current data. The observed fit with square-root form merits further careful investigation given its significance for optimal bidispersity that ensures disorder in granular packs \cite{Luding2001}.

Next, we discuss the stability of these parameters under external perturbation (tapping). As can be seen in Fig.~\ref{fig2}a~\&~b, the values of $\phi~\&~Z$ exhibit no measurable relaxation by tapping, i.e., the initial ($\tau = 0$) and final ($\tau = 10$) configurations fall atop each other; variability being within measurement error. This is in contrast with prior results which demonstrate relaxation in structural parameters (particularly $\phi$) under tapping~\cite{Bandi2013, Iikawa2015}. However, $S$ exhibits measurable relaxation as shown in Fig.~\ref{fig2}c while it follows the same functional form as $\phi~\&~Z$. This is tantamount to force chain orientational order reflecting the pack's metastability at $\tau = 0$ which was not detectable in $\phi~\&~Z$. The metastable relaxation can be detected only by orientational order while configurational disorder remains almost constant. This situation is analogous to liquid crystal nematic phase where the sample is configurationally disordered (liquid) but their orientational order varies independently.

To better understand this metastability and the resulting relaxation under tapping, we plot the evolution of $\phi$, $Z$, and $S$ as a function of the tapping number $\tau$ for crystalline (square and triangular lattice) as well as disordered configurations at $R_S = R_L = 0.5$ (Fig.~\ref{fig2}d, e, \& f). Neither $\phi$ (Fig.~\ref{fig2}d) nor $Z$ (Fig.~\ref{fig2}e) exhibit any dependence on tapping number $\tau$. In Fig.~\ref{fig2}f, the square and triangular lattice configurations exhibit no measurable variation with $\tau$. However, a monotonic decrease was measured in $S$ for disordered configurations ($R_S = R_L = 0.5$). By much more tappings, relaxation might proceed slowly (e.g.~\cite{Rosato2010}). In this study, however, the focus is on parameter sensitivity in particular rather than the slow relaxation. The important fact in the current analysis is that the variation of $S$ is much more significant than $\phi$ and $Z$. The slow relaxation of structure and orientation is under investigation and will be reported in the near future~\cite{Iikawa2016}.

Contact friction is present between disks for all configurations. Metastability arising from friction~\cite{Otsuki2011, Bandi2013} and its resulting protocol dependence~\cite{Inagaki2011, Bandi2013} have been demonstrated in frictional granular packs. Structural relaxation of initial metastable configurations under tapping is therefore to be expected. If friction were the principal source of configurational metastability, it ought to be present and detectable both for crystalline and disordered configurations. However, relaxation in $S$ with tapping is absent for crystalline configurations, but present in disordered configurations, implying $S$ is acutely sensitive to disorder-induced metastable relaxation, as opposed to $\phi~\&~Z$ as normally employed in theory~\cite{Henkes2005}, numerics~\cite{OHern2003} and experiments~\cite{Majmudar2007}. Metastability in the presence of disorder has a rich history in condensed matter systems~\cite{Banerjee1999, Lavallee1991, Henderson1996, Monasson1995}. However, to the best of our knowledge, disorder-induced metastability has not been discussed before in granular media. It is difficult to observe the force chain network in general (thermal and three-dimensional) metastable relaxation phenomena. Rather, the analysis method mentioned in this paper provides a simple way to study the universal physical mechanisms in metastable relaxation by using two-dimensional granular system. In addition, the granular relaxation dynamics offers a fundamental aspect for various natural phenomena such as the development of planetary terrain covered by regolith grains~\cite{Katsuragi2016}.

The tapping-induced decrease of $S$ implies that the force chain structure is statistically reorganized from the vertical to horizontal direction. Whereas one may argue this effect arises from the container's side wall friction, it is unlikely. The values of $\phi$ and $Z$ measured in our experiments (see fig.~\ref{fig2}a and b) are in accord with numerical results of \cite{Silbert2010} where such systematics are absent. Ergo, $\phi(\tau = 0)$ and $Z(\tau = 0)$ have already asymptoted to the Random Loose Packed limit. If side wall friction were dominant, we would expect to see $\phi$ and $Z$ values below the presently observed values. However, finite size effects are certainly manifest and the finite size scaling of the observed behavior is an important problem to be addressed in future.

\begin{figure}
\begin{center}
\includegraphics[width = 3.37 in]{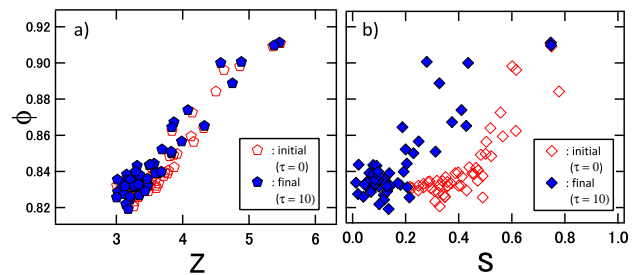}
\end{center}
\caption{(Color online) Relations among packing fraction $\phi$ vs. (a) coordination number $Z$ or (b) force chain orientational order parameter $S$ for all (crystalline as well as disordered) initial ($\tau = 0$; open symbols) and final ($\tau=10$; solid symbols) granular configurations. A linear dependence between $\phi$ and $Z$ can be confirmed. However, no functional form is discernible in $\phi$ vs. $S$.}
\label{fig3}
\end{figure}

Whereas $\phi$ \& $Z$ are structural parameters, $S$ is a topological orientational parameter. Thus, there is no reason to expect a close functional correspondence between these quantities. In fact, structural and orientational orders can be independent in nematic liquid crystals \cite{DeGennes}. However, owing to all three parameters exhibiting a very similar sharp transition in Fig.~\ref{fig2}, we investigated a possible relationship. In Fig.~\ref{fig3}, we plot the $Z$ and $S$ versus $\phi$. 
The identical linear relationship between $Z$ and $\phi$ is observable for both the initial and final states. This tendency is consistent with previous studies~\cite{Rosato2010,Aste2005}. However, $S$ and $\phi$ did not show a systematic relation. This means that while $Z$ and $\phi$ are similar quantities, $S$ is an independent particular quantity to characterize the force chain structure.

In summary, we experimentally studied the order-disorder transition in granular matter by varying bidispersity. We found that a small amount of bidispersity ($R_S \geq 0.1$ or $\mathcal{A}\leq 0.98$) is sufficient to avoid crystallization. In the disordered state, measured parameters ($\phi$, $Z$, and $S$) show almost constant values. Although the structural parameters ($\phi$ and $Z$) are stable under tapping, the force chain orientational order parameter $S$ varies under tapping in disordered states. This implies that $S$ is much more sensitive than traditionally used structural parameters ($\phi$ and $Z$) to detect the granular metastable relaxation that arises from structural disorder alone.

HK was supported by JSPS KAKENHI Grant Number 26610113 and MMB was supported by the Collective Interactions Unit at OIST Graduate University.

\bibliography{all}

\begin{thebibliography}{25}
\bibitem[{not({\natexlab{a}})}]{note2}
\bibinfo{note}{Since the force chain is identified by the thinned structure of
  photoelastic (dark-field) image, the threshold for the force-chain detection
  comes from the sensitivity limit of photoelastic materials.}
\bibitem[{not({\natexlab{b}})}]{note1}
\bibinfo{note}{Contact is defined by the threshold $0.99(R_1+R_2) < D \le
  1.02(R_1+R_2)$; $R_1 \& R_2$ are radii of two neighboring disks and $D$ the
  measured distance between their centers.}
\end{thebibliography}

\end{document}